\documentclass[manuscript]{acmart}
\AtBeginDocument{%
  }

\usepackage{xcolor}

\begin{document}

\title{Supporting Data-Frame Dynamics in AI-assisted Decision Making}

\begingroup
\renewcommand\thefootnote{}\footnote{
 This paper was presented at the 2025 ACM Workshop on Human-AI Interaction for Augmented Reasoning (AIREASONING-2025-01). This is the authors’ version for arXiv.}
\endgroup

\author{Chengbo Zheng}
\affiliation{%
 \institution{University of Queensland}
 \city{Brisbane}
 \state{Queensland}
 \country{Australia}}

 \author{Tim Miller}
\affiliation{%
 \institution{University of Queensland}
 \city{Brisbane}
 \state{Queensland}
 \country{Australia}}

 \author{Alina Bialkowski}
\affiliation{%
 \institution{University of Queensland}
 \city{Brisbane}
 \state{Queensland}
 \country{Australia}}

 \author{H Peter Soyer}
\affiliation{%
 \institution{University of Queensland}
 \city{Brisbane}
 \state{Queensland}
 \country{Australia}}

 \author{Monika Janda}
\affiliation{%
 \institution{University of Queensland}
 \city{Brisbane}
 \state{Queensland}
 \country{Australia}}
\renewcommand{\shortauthors}{Zheng et al.}

\begin{abstract}
High stakes decision making often requires a continuous interplay between evolving evidence and shifting hypotheses, a dynamic that is not well supported by current AI decision support systems. 
In this paper, we introduce a mixed-initiative framework for AI assisted decision making that is grounded in the data frame theory of sensemaking and the evaluative AI paradigm. 
Our approach enables both humans and AI to collaboratively construct, validate, and adapt hypotheses. 
We demonstrate our framework with an AI assisted skin cancer diagnosis prototype that leverages a concept bottleneck model to facilitate interpretable interactions and dynamic updates to diagnostic hypotheses.
\end{abstract}

\keywords{AI-assisted Decision Making, Explainable AI, Human-AI Collaboration, Sensemaking}

\maketitle

\section{Introduction}
Decision-making in high-stakes scenarios is inherently dynamic. 
Consider the clinical context: a physician may initially identify a few key features in a patient's profile that narrow the diagnostic options. 
Yet rather than committing immediately to these options, the doctor is likely to continuously integrate additional data, such as subtle clinical signs or evolving test results, to either reinforce the initial hypothesis or necessitate a shift to an alternative hypothesis.
This iterative process is at the essence of the \textbf{data-frame theory} of sensemaking, which is developed through field observations of people decision making in complex and real-world environments \citep{klein2007data}.
In the data-frame theory , data refers to the signals extracted from the contexts (e.g., a dermatologist's observation of a lesion's dark color or irregular shape), while the frame represents the model or hypothesis that structure those observations (e.g., the lesion is melanoma). At its essence, the data-frame theory describes how people iteratively fit data into frames and fit frames into data, ultimately achieving their congruence.

\begin{figure}
    \centering
    \includegraphics[width=0.9\linewidth]{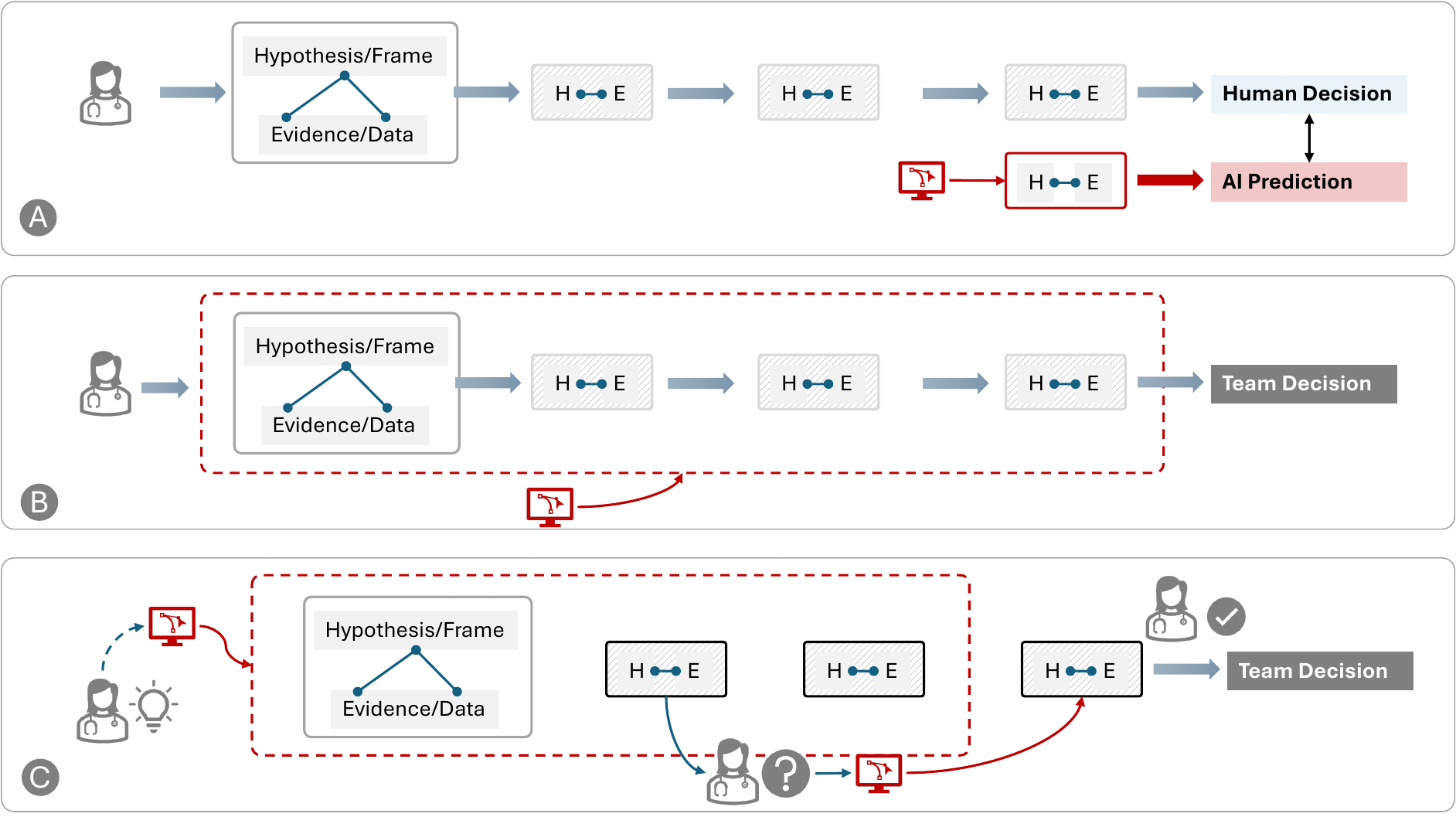}
    \caption{(A) Typical AI decision support: The human’s decision and the AI’s prediction can be at odds, requiring the human to interpret the AI’s output and resolve conflicts independently; (B) The evaluative AI framework: The AI generates hypotheses and corresponding evidence to facilitate sensemaking. However, how to generate the support to well synchronize with human reasoning is unclear; (C) Instantiate evaluative AI with mixed-initiative sensemaking: Humans can propose reasoning, and the AI actively integrates these updates to offer support, enabling collaborative, team-based decision making.}
    \label{fig:idea}
\end{figure}

In various domains~\cite{militello2018understanding, hudson2017expert, moore2011data}, the dynamics of data and frames have been well documented, yet many AI-driven decision support systems do not adequately incorporate these interactions into their design. 
Most current systems focus on providing decision recommendations (i.e., frames) and possibly explanations linking the data to those frames~\cite{bansal2021does, ma2023should, yin2019understanding}. 
By fixating on a single data-frame structure, such systems often overlook alternative possibilities and stray from the cognitive needs of human decision makers. 
When a user's hypothesis diverges from AI’s, the human-AI relationship can shift from collaboration to competition (\autoref{fig:idea} (A)), increasing the cognitive burden on the user or leading to overreliance if they abandon their own perspective.

Recently, \citet{miller2023explainable} introduced ``evaluative AI,'' a conceptual framework for AI-assisted decision making rooted in abductive reasoning and data-frame theory. 
The framework advocates for AI systems that, rather than merely offering recommendations, empower decision makers to formulate multiple hypotheses (i.e., frames) and evaluate them against data (\autoref{fig:idea} (B)), thereby aligning the reasoning processes of both human and machine.
While this approach lays a conceptual framework for collaborative decision support, it leaves several key implementation challenges unresolved. 
For example, what specific support should the AI system provide to facilitate the data-frame sensemaking process? 
If users find the system's proposed data-frame questionable, how can they intervene in the process, and how should AI adapt to offer continuous, context-aware support?

In this paper, we provide a method that instantiates  \citet{miller2023explainable}'s evaluative AI via a \textit{mixed-initiative human-AI collaborative sensemaking} workflow that accounts for data-frame dynamics (\autoref{fig:idea} (C)). 
By ``mixed-initiative''~\cite{allen1999mixed}, we refer to a process where both the human and the AI can contribute to the data-frame dynamics, thereby jointly shaping their understanding of the task. 
By formalizing the sensemaking process, we propose the decision support system should clearly communicate AI-driven processes of evidence extraction, hypothesis retrieval, and hypothesis scoring.
On the other hand, human decision makers should be empowered to easily express their own reasoning across these three critical sensemaking functions.
Human inputs should also directly influence AI reasoning, progressively aligning the human-AI team's collaborative sensemaking.
We see two key advantages in this method. 
First, users are not compelled to make a binary judgment on AI's logic (i.e., accept or reject), potentially reducing overreliance and the cognitive load. 
Second, AI can continue offering decision support over time grounded on the users' just-in-time reasoning, which might facilitate engagement, thus benefit skill learning over time.

In the following sections, we first present our vision for human-AI collaborative sensemaking based on the data-frame theory. 
Then, to illustrate this concept, we introduce our AI-assisted skin cancer diagnosis prototype.

\section{Support Data-Frame Dynamics through Human-AI Collaboration}
\label{sec:interaction}
While there are other theoretical models characterizing the sensemaking process~\cite{russell1993cost, pirolli2005sensemaking, weick1995sensemaking}, \citet{klein2007data}'s data-frame theory is developed with a focus on high-stakes decision making~\cite{klein2005recognition}. 

Here, we formalize \citet{klein2007data}'s data-frame theory.
In order to fit the context of decision making -- the focus of this paper -- we adopt the terminology of \emph{evidence} for data and \emph{hypotheses} for frames, following \citet{miller2023explainable}. 

Let $I$ represent the set of raw information available, $E_t$ denote the set of evidence derived from this raw information at time step $t$, and $\mathcal{H}_t$ indicate the set of hypotheses under consideration at time $t$. 
The sensemaking process begins by transforming raw information into evidence via an extraction function $f_t: I_t \rightarrow E_t$, such that $E_t = f_t(I_t)$. Given the current evidence $E_t$, a retrieval function $r_t: E_t \rightarrow 2^{\mathcal{H}}$ selects a subset of relevant hypotheses, denoted as $\mathcal{H}_t = r_t(E_t)$. Each hypothesis $H \in \mathcal{H}_t$ is evaluated against the evidence using a scoring function $s: E_t \times \mathcal{H}_t \rightarrow [0,1]$, resulting in a score $s(E_t,H)$ that indicates how well each hypothesis explains the evidence. 

If at time $t$ the highest-scoring hypothesis does not meet or exceed an acceptance threshold $\delta$, i.e., if $\max_{H \in \mathcal{H}_t} s(E_t,H) < \delta$, one of, or both the two possible iterative refinement actions can occur at the next time step $t+1$: 1) additional evidence is generated from raw information, $E_{t+1} = E_t \cup f_{t+1}(I)$; 2) additional hypotheses are retrieved to expand the candidate set, $\mathcal{H}_{t+1} = \mathcal{H}_t \cup r_{t+1}(E_{t})$ or $\mathcal{H}_{t+1} = \mathcal{H}_t \cup r_{t+1}(E_{t+1})$. 
This iterative updating continues dynamically over multiple time steps until at least one hypothesis exceeds the threshold $\delta$, formally expressed as:
\[
H_t^* = \arg\max_{H \in \mathcal{H}_t} s(E_t,H), \quad \text{subject to} \quad s(E_t,H_t^*) \geq \delta.
\]


To support the dynamics of the sensemaking, we propose that the AI decision support should assist in evidence extraction $f_t$, hypotheses retrieval $r_t$, and scoring of hypotheses versus evidence $s_t$.
Besides, the human decision makers might have different functions, thus user-initiated interactions should be supported in intervening the above three functions.
Such an intervention should also introduce updates to AI's reasoning by introducing $f_{t+1}$, $r_{t+1}$ and $s_{t+1}$.

More specifically, first, for \textbf{the evidence extraction function} $f_t$, AI could display the extracted evidence in an interpretable format and allow users to annotate, modify, or augment this evidence. 
Feedback from the human decision maker introduces $f_{t+1}$, $E_{t+1}$ as well as $\mathcal{H}_{t+1}$ from $r_t(E_{t+1})$. 
The latter two updates should be clearly communicated with the users to lay down the common ground.
Second, for \textbf{the hypotheses retrieval function} $r_t$, AI should enable users to review the candidate hypotheses, suggest additional ones, or exclude those that appear irrelevant. 
Third, for \textbf{the hypotheses scoring function} $s(E_t,H)$, AI should clear communicate its weightings of the evidences that influence the evaluation. Users should be empowered to discuss new weightings to shape the team scoring of the hypotheses.

\section{Example: AI-assisted Skin Cancer Diagnosis}
\begin{figure}
    \centering
    \includegraphics[width=\linewidth]{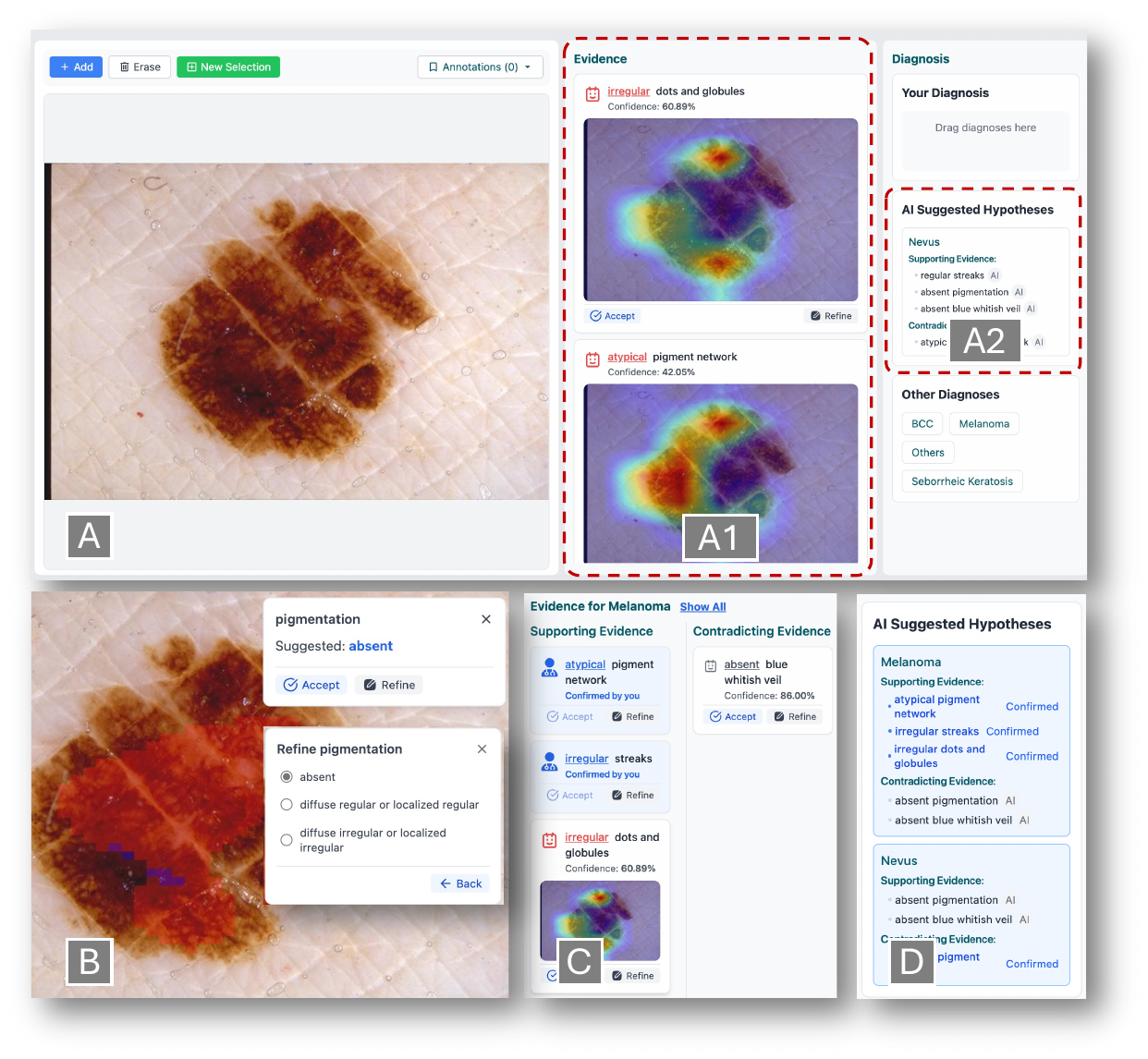}
    \caption{An example system that implements the mixed-initiative sensemaking support for skin cancer diagnosis (the image is from the derm7pt dataset~\cite{kawahara2018seven}).}
    \label{fig:system}
\end{figure}

In this section, we present an AI-assisted skin cancer diagnosis prototype (\autoref{fig:system}) to examplify the implementation of the three key functions for supporting data-frame dynamics outlined in \autoref{sec:interaction}. 
The interface comprises three primary views (from left to right in \autoref{fig:system}~(A)):
\begin{itemize}
\item \textbf{Image view:} displays the dermoscopy image.
\item \textbf{Evidence view:} lists evidences extracted by the AI or provided by the user.
\item \textbf{Diagnosis view:} allows the user to manage hypotheses and finalize the diagnosis.
\end{itemize}

We trained a concept bottleneck model (CBM)\cite{koh2020concept} on the derm7pt dataset\cite{kawahara2018seven} to power the prototype. 
CBM is an interpretable deep learning approach that first maps image representations to task-relevant concepts, then makes final predictions based on these concepts. 
The seven dermoscopy concept annotations (e.g., pigment network, streaks) in the derm7pt dataset make it well-suited for training a CBM. 
These concepts can also be viewed as critical clinical evidences that corresponds to the data in the \citet{klein2007data}'s sensemaking theory.
Moreover, CBM enables users to edit the predicted concepts, thus influencing the final diagnostic outcome~\cite{koh2020concept}.

The prototype implements the three key functions mentioned in \autoref{sec:interaction} to facilitate the sensemaking process in dermatological diagnosis:

\begin{itemize}
    \item \textbf{Evidence Extraction}.
    Initially, the system presents AI-extracted evidence in the \textit{Evidence View}, utilizing a heatmap to visually highlight how the evidence was derived from the raw input (computed using GradCam~\cite{selvaraju2017grad} based on concept predictions within CBM).
    Upon first viewing the raw input, human decision-makers might not check the AI-extracted evidence but quickly anchor on specific pieces of evidence (key anchoring data, as discussed by \citet{klein2007data}).
    Besides, they might find discrepancies between AI-generated evidence and their own understanding, or determine that evidence updates are necessary when evaluating hypotheses against existing evidence.
    To address these scenarios, the system supports user-initiated interventions for evidence extraction. 
    If a user considers certain evidence inaccurate or incomplete, they can refine it using the ``Refine'' button (see \autoref{fig:system}(B) and \autoref{fig:system}~(C)). 
    Such refinements update the corresponding concept predictions in the CBM, consequently adjusting AI-generated hypotheses.
    Additionally, if users identify suspicious regions, they can annotate these regions (purple patches in \autoref{fig:system}(B)). 
    Although users can manually label these annotations, the system proactively proposes potential evidence related to the user's annotation (illustrated as red patches in \autoref{fig:system}(B) and within pop-up windows). 
    Users may accept or further modify these AI-proposed evidences

    \item \textbf{Hypotheses Retrieval}.
    The system automatically suggests hypotheses based on the current set of evidence (\autoref{fig:system}(A2)). 
    The hypotheses set is generated through conformal prediction~\cite{angelopoulos2020uncertainty}. 
    This set is also dynamic.
    When newly confirmed evidence introduces additional plausible hypotheses, the system recommends these new options (e.g., ``Melanoma'' appearing in \autoref{fig:system}(D) after user refinement of evidence).
    Additionally, users can introduce new hypotheses if the AI suggestions do not fully align with their reasoning or if updates required as no hypothesis satisfies the acceptance threshold after the scoring.
    This is done by dragging options from the ``Other Diagnosis'' section to the ``Your Diagnosis'' section within the Diagnosis View.
    
    \item \textbf{Hypotheses Scoring}.
    Each hypothesis is accompanied by clearly indicated supporting and contradicting evidence (\autoref{fig:system}~(D)), generated based on weights from the final prediction layer in the CBM, using concept predictions as inputs.
    Confirmed evidence by users and AI-generated, user-unverified evidence are visually distinguished in the interface (e.g., blue points for confirmed evidence and grey points for unverified evidence in \autoref{fig:system}~(D)).
    Users can also explore the scoring of individual hypotheses in greater depth by selecting a specific hypothesis within the Evidence View (\autoref{fig:system}~(C)).
    They can remove evidences from the supporting and contradicting groups, or move evidence between groups (though such cases are expected to be rare in skin cancer diagnosis as general consensus on the weights of the clinical concepts exists~\cite{kawahara2018seven}).
    
\end{itemize}






\section{Conclusion}
In this work, we instantiates evaluative AI~\cite{miller2023explainable} by introducing a mixed-initiative interaction framework that bridges human insight and machine reasoning for enhanced decision making. 
Grounded in \citet{klein2007data}'s data-frame theory, our approach suggests supporting three sensemaking functions -- evidence extraction, hypotheses retrieval, and hypotheses scoring -- that empower human decision makers to collaborate with AI. 
We introduce an AI-assisted skin cancer diagnosis prototype as a proof-of-concept of the interaction framework.
Further study with dermatologists on the prototype will be conducted to evaluate the effectiveness and tradeoffs of the interaction framework.


\bibliographystyle{ACM-Reference-Format}
\bibliography{sample-base}


\end{document}